\documentclass[twocolumns,10pt,twoside]{IEEEtran}
\usepackage{epsfig,amsfonts}
\usepackage[nolist]{acronym}
\usepackage{graphicx,cite,amssymb,amsmath}
\usepackage[usenames,dvipsnames]{color}
\usepackage{color}
\usepackage{psfrag}
\usepackage{amsbsy}
\usepackage{subfigure, amsbsy}
\usepackage{perso}
\usepackage{tikz}
\usetikzlibrary{shapes,arrows}

\IEEEoverridecommandlockouts

\tikzstyle{block} = [draw, fill=white!20, rectangle, minimum
height=3em, minimum width=7em, text width=6em, text centered]
\tikzstyle{point} = [draw, fill=black!20, circle]
\tikzstyle{input}=[coordinate] \tikzstyle{output} = [coordinate]
\tikzstyle{left_of_block} = [coordinate] \tikzstyle{left_of_LRFC}
=[coordinate] \tikzstyle{r_input} =[coordinate]

\begin{document}
\begin{acronym}
\acro{LT}{Luby-transform}
\acro{ML}{maximum-likelihood}\acro{FEC}{forward error
correction}\acro{RS}{Reed-Solomon}\acro{ARQ}{Automatic
Retransmission Query}\acro{LRFC}{linear random fountain
code}\acro{MDS}{maximum distance separable} \acro{GE}{Gaussian
elimination}\acro{BP}{belief propagation}\acro{SPC}{single
parity-check}\acro{GRS}{generalized Reed-Solomon} \acro{WE}{weight-enumerator} \acro{WEF}{weight-enumerator function}
\acro{C-OWEF}{conditional output-weight enumerator function}\acro{BEC}{binary erasure channel}
\end{acronym}

\title{Parallel Concatenation of Non-Binary Linear Random Fountain Codes with Maximum Distance Separable Codes}
\author{
Francisco L\'azaro, Giuliano Garrammone, Gianluigi Liva
\thanks{This work was presented in part at IEEE International Conference on Communications 2011, Kyoto. Francisco L\'azaro, Giuliano Garrammone and Gianluigi Liva are with the Institute of Communications and
Navigation, German Aerospace Center (DLR), Oberpfaffenhofen, 82234
Wessling, Germany.
Email:\{Francisco.LazaroBlasco,Giuliano.Garrammone,\newline Gianluigi.Liva\}@dlr.de.

This work has been accepted for publication in IEEE Transactions on Communications:
Digital Object Identifier 10.1109/TCOMM.2013.09.120834

\copyright 2013 IEEE. Personal use of this material is permitted. Permission
from IEEE must be obtained for all other uses, in any current or future media, including
reprinting /republishing this material for advertising or promotional purposes, creating new
collective works, for resale or redistribution to servers or lists, or reuse of any copyrighted
component of this work in other works}} \maketitle

\begin{abstract}
The performance and the decoding complexity of a novel coding scheme based on the concatenation of maximum
distance separable (MDS) codes and \acl{LRFC}s  are investigated.  Differently from Raptor codes
(which are based on a serial concatenation of a high-rate outer
block code and an inner Luby-transform code), the proposed coding
scheme can be seen as a parallel concatenation of a MDS code and a \acl{LRFC},
 both operating on the same finite
field. Upper and lower bounds on the decoding failure probability
under maximum-likelihood (ML) decoding are developed. It is shown
how, for example, the concatenation of a $(15,10)$ Reed-Solomon (RS)
code and a \acl{LRFC} over a finite field of order $16$, $\mathbb {F}_{16}$,
brings to a decoding failure probability $4$ orders of magnitude
lower than the one of a \acl{LRFC} for the same receiver overhead in
a channel with a erasure probability of
$\epsilon=5\cdot10^{-2}$. It is illustrated how the performance of
the novel scheme approaches that of an idealized
fountain code for higher-order fields and moderate erasure probabilities. An efficient decoding algorithm is developed for the case of a (generalized) RS code.
\end{abstract}


{\pagestyle{plain} \pagenumbering{arabic}}


\section{Introduction}\label{sec:Intro}

Efficient reliable multicasting/broadcasting techniques
have been
 investigated during the past thirty years
\cite{Metzer84:retransmission} and especially during the past decade
\cite{byers02:fountain,luby02:LT,shokrollahi06:raptor,Medard08:ARQ,Liva10:fountain,liva2010carq,blasco2011concatenation,schotsch2011performance,Vary2011:Allerton}.
Perhaps, the most successful approach to reliable multicast deals
with the so-called fountain codes \cite{byers02:fountain}. Consider
the case where a sender (or source) needs to deliver a source block (e.g., a file) to a set
of $N$ receivers. Consider furthermore the case where receivers are
affected by packet losses. In this scenario, the usage of an
\ac{ARQ} protocol can result in large inefficiencies, since receivers
may loose different packets, and hence a large number of
retransmissions would crowd the downlink channel.  When a fountain
code is used, the source block is split in a set of $k$ source
packets, which we will denote as source symbols. The sender computes linear
combinations (also referred to as fountain coded packets, or output symbols) of the $k$ source packets and broadcasts them through
the communication medium. After receiving $k$ fountain coded
packets, the receivers can try to recover the source packets. {In case of decoding failure, they will} try again to decode
{after receiving} additional packets.
The efficiency of a fountain code deals with the amount of packets
 that a receiver needs to collect for
recovering the source {block}. An \emph{idealized} fountain code would
allow the recovery with a failure probability $P_f=0$ from any
set of $k$ received packets. Actual fountain decoders need in
general to receive a larger amount of packets, $m=k+\delta$, for
succeeding in the recovery. Commonly, $\delta$ is referred to as (receiver)
\emph{overhead} of the fountain code, and is used to measure its
efficiency.
The first class of practical fountain codes are \ac{LT} codes
\cite{luby02:LT}. Among them, random \ac{LT} codes or \acp{LRFC}
{\cite{shokrollahi06:raptor,Medard08:ARQ}} deserve a particular attention due to
their excellent performance and to the {relatively simple} performance
model. Under \ac{ML} decoding,  the failure probability of a binary
\ac{LRFC} {\cite{shokrollahi06:raptor,Medard08:ARQ}} can be accurately
modeled as $P_f\sim 2^{-\delta}$ for $\delta\geq2$. It can be proved
that $P_f$ is actually always upper bounded by $2^{-\delta}$
{\cite{berlekamp:bound,shokrollahi06:raptor,Medard08:ARQ}}. In
\cite{Liva10:fountain,schotsch2011performance} it was shown that
this expression is still accurate for fountain codes based on sparse
matrices (e.g., Raptor codes {\cite{shokrollahi06:raptor}})
under \ac{ML} decoding. In \cite{Liva10:fountain}, the
performance achievable by performing linear combinations of packets
on finite fields of order {larger} than $2$ ($\mathbb {F}_q$,
$q>2$) was analyzed. For a \ac{LRFC} over $\mathbb {F}_q$,
the failure probability under \ac{ML} decoding is bounded as
\cite{Liva10:fountain}
\begin{equation}\label{eq:tightbounds}
q^{-\delta-1}\leq P_f(\delta,q) < \frac{1}{q-1}q^{-\delta}
\end{equation}
where both bounds are tight already for $q=2$, and become tighter for
increasing $q$. The improvement in efficiency obtained by fountain
codes operating on fields of order larger than $2$ has been analyzed
in \cite{Liva10:fountain,Vary2011:Allerton} and has led to recent
standardization activities \cite{lubyraptorq}. In
\cite{Liva10:fountain,Vary2011:Allerton} it was also shown that
non-binary Raptor and \ac{LT} codes can in fact tightly approach the
bounds \eqref{eq:tightbounds} down to moderate error rates under
\ac{ML} decoding. Thus, \eqref{eq:tightbounds} can be successfully
used to model the performance of common classes of fountain codes.
The result is remarkable considering that for Raptor codes, under \ac{BP} decoding, both the encoding and decoding costs\footnote{The
cost is defined as the number of arithmetic field operations divided by
the number of source symbols, $k$.} are
 $\mathcal O(\log(1/\varepsilon) )$ {\cite[Theorem 5]{shokrollahi06:raptor}},
being $\varepsilon=\delta/k$ the overhead (normalized to $k$) needed
to recover the source symbols with a high probability. For a
\ac{LRFC} the encoding cost is $\mathcal O(k)$ and the decoding cost
is $\mathcal O(k^2)$, and thus it does not scale favorably with the
source block size. However,  \ac{BP} decoding is scarcely used in
practical Raptor decoder implementations \cite{MBMS05:raptor} due
its poor performance with source block lengths  of practical interest ($k$
up to few thousands symbols). Efficient \ac{ML} decoding algorithms
based on \ac{GE} are usually adopted
\cite{lamacchia91:solving,studio3:RichardsonEncoding,miller04:bec,shokrollahi2005systems,MBMS05:raptor,paolini12:TCOM},
for which the decoding cost is $\mathcal O(k^2)$, though  the
fraction of symbols that are recovered with quadratic cost can be
kept remarkably small. Similarly, in the short source block length
regime, the application of \acp{LRFC} under \ac{GE} decoding is
usually considered practical
\cite{Liva10:fountain,Vary2011:Allerton}.

In this paper, we introduce and analyze  a further improvement of
the approach proposed in \cite{Liva10:fountain,Vary2011:Allerton}
{to design} fountain codes with good performance for short block
lengths. More specifically, a $(n,k)$ \ac{MDS} code is introduced in
parallel concatenation with the \ac{LRFC}. By doing that, the first
$n$ output symbols {are the codeword symbols of the \ac{MDS} code.}\footnote{This represents a crucial difference with Raptor
codes, for which the output of the precode is further encoded by a
\ac{LT} Code. Hence the first $n$ output symbols of a Raptor encoder
do not coincide with the output of the precode.}
We will assume that the
\ac{MDS} linear block code is constructed on the same field $\mathbb
{F}_q$ {as} the fountain code.
{A related rate-less construction was proposed in \cite{kasai}, where a mother non-binary low-density parity-check code was modified by replicating the codeword symbols (prior multiplication by a non-zero field element) and thus by (arbitrarily) lowering the code rate. In our work, the mother code is a \ac{MDS} code, while additional redundant symbols are produced by a linear random fountain encoder.}
For the proposed scheme, we illustrate how the performance of
\acp{LRFC} in terms of probability of decoding failure can be
remarkably improved thanks to the concatenation, especially for low to moderate
packet loss probabilities. Tight bounds on the decoding failure
probability vs. overhead {are} derived under the assumption of
\ac{ML} decoding.  The accuracy of the bounds is  confirmed through
simulations. An efficient \ac{ML} decoding algorithm is presented for
the case where a (generalized) \ac{RS} is used in the concatenation.
An analysis for the general case where the \ac{MDS} code is replaced by any arbitrary
linear block code, in a finite rate regime, is provided in the Appendix.

The paper is organized as follows. In Section
\ref{sec:concatenation} the proposed concatenated scheme is
introduced. Section \ref{sec:eff_decoding} provides an efficient {\ac{ML}} decoding algorithm. In Section \ref{sec:bounds} the performance {is analyzed and tight bounds on the decoding failure probability are derived}, while numerical results are presented in Section
\ref{sec:results}. Conclusions follow in Section \ref{sec:conc}.


\section{Concatenation of Block Codes with {Linear Random} Fountain
Codes}\label{sec:concatenation}

We define the
source block $\mathbf{u}=(u_1, u_2, \ldots, u_k)$ as a vector of source
symbols belonging to a finite field of order $q$, i.e.,
$\mathbf{u}\in \mathbb {F}_q^k$. In the proposed approach, the
source block is first encoded via a $(n,k)$ linear block
code $\mathcal{C}'$ over $\mathbb {F}_q$ with generator matrix
$\mathbf{G}'$. The encoded block is hence
given by
$\mathbf{c}'=\mathbf{u}\mathbf{G}'=(c'_1,c'_2,\ldots,c'_n)$. Additional
redundancy symbols can be obtained by computing {linear random}
combinations of the $k$ source symbols as
\begin{equation}
c_i=c_{i-n}''=\sum_{j=1}^{k}g_{j,i}u_j, \qquad
i=n+1,\ldots, l
\label{eq:encoding}
\end{equation}
where the coefficients $g_{j,i}$ in \eqref{eq:encoding} are picked from $\mathbb {F}_q$
with a uniform probability.

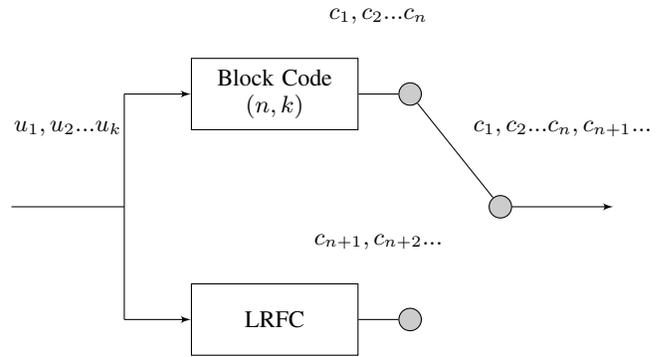
\begin {figure}[h]
\begin{center}
{\small
\begin{tikzpicture}[auto, node distance=1.3cm, label distance=6mm, >=latex']
    \node [input, name=input] {};
    \node [r_input, right of=input, node distance = 1.5cm](r_input) {};
    \node [left_of_block, above of =r_input, node distance = 1.5 cm] (left_of_block) {};
    \node [left_of_LRFC, below of =r_input, node distance = 1.5 cm] (left_of_LRFC) {};
    \node [block, right of=left_of_block,  node distance = 2 cm] (blockcode) {Block Code  $(n,k)$};
    \node [block, right of=left_of_LRFC, , node distance = 2 cm] (LRFC) {LRFC};
    \node [point, right of=blockcode, node distance = 1.8 cm] (blockcode_point){};
    \node [point, right of=LRFC, node distance = 1.8 cm] (LRFC_point) {};
    \node [point, right of=input, node distance = 6.5cm] (out_point) {};
    \node [output, right of=out_point, node distance = 1.5 cm] (output) {};
    \draw [-]  (input) -- node [label=above:{$u_1,u_2...u_k$}] { } (r_input);
    \draw [-]  (r_input) -- node  { } (left_of_block);
    \draw [-]  (r_input) -- node { } (left_of_LRFC);
    \draw [->] (left_of_block) -- node {} (blockcode);
    \draw [->] (left_of_LRFC) -- node {} (LRFC);
    \draw [-] (blockcode) -- node
    [label=above:{$c_1,c_2...c_n$}] {} (blockcode_point);
    \draw [-] (LRFC) -- node
    [ label=above:{$c_{n+1},c_{n+2}...$}] {} (LRFC_point);
    \draw [->] (out_point) -- node
    [label=above:{$c_1,c_2...c_n,c_{n+1}...$}] {} (output);
    \draw [-] (blockcode_point) -- node{} (out_point);
\end{tikzpicture}}
\caption{Fountain coding scheme seen as a parallel concatenation of
a $(n,k)$ linear block code and a linear random fountain
code.}\label{fig:par}
\end{center}
\end {figure}

{The encoded sequence is thus}
 $\mathbf{c}=(\mathbf{c}'|\mathbf{c}'')$. The
generator matrix of the concatenated code has the form
\begin{equation}
\mathbf{G}=
\underbrace{\left(\begin{array}{cccc}
  g_{1,1} & g_{1,2} & \ldots & g_{1,n} \\
  g_{2,1} & g_{2,2} & \ldots & g_{2,n} \\
  \vdots  & \vdots  & \ddots & \vdots  \\
  g_{k,1} & g_{k,2} & \ldots & g_{k,n}
\end{array}\right|}_{\mathbf{G}'}  \underbrace{\left|\begin{array}{cccc}
  g_{1,n+1} & g_{1,n+2} & \ldots  & g_{1,l} \\
  g_{2,n+1} & g_{2,n+2} & \ldots  & g_{2,l} \\
  \vdots    & \vdots    & \ddots  & \vdots  \\
  g_{k,n+1} & g_{k,n+2} & \ldots  & g_{k,l}
\end{array}\right)}_{\mathbf{G}''}
\end{equation}
where $\mathbf{G}''$ is the generator matrix of the \ac{LRFC}. Note
that, being the \ac{LRFC} rate-less, the number $l$ of columns of
$\mathbf{G}$ can grow indefinitely. The encoder can be
seen hence as a parallel concatenation of the linear block code
$\mathcal C '$ and of a \ac{LRFC} (Fig. \ref{fig:par})  {and the encoded sequence can be written as
$\mathbf{c}=\mathbf{u}\mathbf{G}=(c_1,c_2,\ldots,c_l)$.} {The proposed construction allows generating infinitely many redundancy symbols. Thus, the encoder may be seen as a modified fountain encoder, whose first $n$ output symbols $(c_1,c_2,\ldots,c_n)$ correspond to the codeword output by the encoder of $\mathcal{C}'$, whereas the following $l-n$ symbols are the output of the \ac{LRFC} encoder.}
%

\section{Efficient Decoding}\label{sec:eff_decoding}
{We consider a multicast setting, where a number of receivers try to retrieve the source block from the respectively-received output symbols. In this context, the decoder behaves as for a conventional fountain decoder. At each receiver, the correctly-received output symbols are forwarded to the decoder. As soon as  $k$ output symbols are collected, a decoding attempt is performed. If the decoding is not successful, further output symbols are collected. Whenever an additional output symbol is received, another decoding attempt is performed. In case of successful decoding, the receiver acknowledges the correct reception. The overall number of symbols collected at a receiver is denoted by $m=k+\delta$ (recall that $\delta$ is referred to as the overhead). On the encoder side, as soon as a target success rate  among the receivers is attained, encoding stops. Note that at each receiver,  the $m$ output symbols that are collected may belong to
\begin{itemize}
\item[i)] the output of the $\mathcal C'$ encoder only,
\item[ii)] the output of the \ac{LRFC} encoder only,
\item[iii)] both the outputs of the  $\mathcal C'$ encoder and the \ac{LRFC} encoder.
 \end{itemize}
 While in the third case there is no different with respect to a classical LRFC case, in the other two cases the structure of the $\mathcal C'$ generator matrix can be exploited to reduce the decoding complexity, as we will see next. Furthermore, when the channel erasure probability is sufficiently low, the event i) may dominate, leading to a remarkable improvement in the decoding failure probability. In this sense, the proposed scheme provides the same performance of a (universal) \ac{LRFC} at high channel erasure probabilities, whereas it will enjoy a boost in the efficiency when the channel erasure probability is low.}
We denote
by $\msr J=\{j_1, j_2, \ldots, j_{m}\}$ the set of the indexes on
the symbols of $\mathbf{c}$ that have been {collected by a specific receiver}. The received
vector {$\mathbf{y}$} is hence given by
\[
\mathbf{y}=(y_1, y_2, \ldots,
y_{m})=(c_{j_1},c_{j_2},\ldots,c_{j_{m}})
\]
and it can be related to the source block $\mathbf{u}$ as
{$
\mathbf{y}=\mathbf{u}\tilde{\mathbf{G}}$.}
 Here,
$\tilde{\mathbf{G}}$ denotes the $k\times m$ matrix made by the
columns of $\mathbf{G}$ with indexes in $\msr J$, i.e.,
\[
\tilde{\mathbf{G}}= \left(\begin{array}{cccc}
  g_{1,j_1} & g_{1,j_2} & \ldots & g_{1,j_{m}} \\
  g_{2,j_1} & g_{2,j_2} & \ldots & g_{2,j_{m}} \\
  \vdots  & \vdots  & \ddots & \vdots  \\
  g_{k,j_1} & g_{k,j_2} & \ldots & g_{k,j_{m}}
\end{array}\right).
\]
The recovery of $\mathbf{u}$ reduces to solving the system  of
$m=k+\delta$ linear equations in $k$ unknowns
\begin{equation}
\tilde{\mathbf{G}}^T\mathbf{u}^T=\mathbf{y}^T.\label{eq:solve}
\end{equation}
{The solution of \eqref{eq:solve} can be obtained (e.g., via Gaussian
elimination)}  if and only if $\textrm{rank}
(\tilde{\mathbf{G}})=k$.

Assuming $\mathcal C'$ being \ac{MDS}, the system
is solvable with probability $1$ if, among the $m$ received symbols,
at least $k$ have indexes in $\{1, 2, \ldots, n\}$, i.e., if at least
$m'\geq k$ symbols produced by the linear block encoder have been
received. {Let us consider the less trivial case where $m'<k$
among the $m$ received symbols have indexes in $\{1, 2, \ldots,
n\}$. We can partition $\tilde{\mathbf{G}}^T$ as
\begin{equation}
\tilde{\mathbf{G}}^T=\left(\begin{array}{c} \tilde{\mathbf{G}}'^T\\
\tilde{\mathbf{G}}''^T \end{array}\right)=\left(\begin{array}{cccc}
  g_{1,j_1} & g_{2,j_1} & \ldots & g_{k,j_1} \\
  g_{1,j_2} & g_{2,j_2} & \ldots & g_{k,j_2} \\
  \vdots  & \vdots  & \ddots & \vdots  \\
  g_{1,j_{m'}} & g_{2,j_{m'}} & \ldots & g_{k,j_{m'}}\\ \hline
  g_{1,j_{m'+1}} & g_{2,j_{m'+1}} & \ldots & g_{k,j_{m'+1}} \\
  g_{1,j_{m'+2}} & g_{2,j_{m'+2}} & \ldots & g_{k,j_{m'+2}} \\
  \vdots  & \vdots  & \ddots & \vdots  \\
  g_{1,j_{m}} & g_{2,j_{m}} & \ldots & g_{k,j_{m}}
\end{array}\right). \label{eq:G_partition}
\end{equation}
The \ac{MDS} property of $\mathcal{C}'$ assures that $\textrm{rank}
(\tilde{\mathbf{G}}')=m'$, i.e., the first $m'$ rows of
$\tilde{\mathbf{G}}^T$ are linearly independent. Note that the
$m''\times k$ matrix $\tilde{\mathbf{G}}''^T $ (with $m''=m-m'$) {can be modeled as
a random matrix whose elements are uniformly distributed  in
$\mathbb F _q$.} It follows that the {matrix in}
\eqref{eq:G_partition} can be put (via column permutations over
$\tilde{\mathbf{G}}^T$ and row permutations/combinations over
$\tilde{\mathbf{G}}'^T$) in the form
\begin{equation}
\hat{\mathbf{G}}^T=\left(\begin{array}{ccc} \mathbf{I} & \vline &
\mathbf{A} \\\hline
\mathbf{0} & \vline & \mathbf{B}\\
\end{array}\right), \label{eq:G_partition_manipulation}
\end{equation}
where $\mathbf I$ is the $m' \times m'$ identity matrix,
$\mathbf{0}$ is a $m'' \times m'$ all-$0$ matrix, and  $\mathbf{A}$,
$\mathbf{B}$ have respective sizes $m' \times (k-m')$ and $m''
\times (k-m')$. Note that the lower part of $\hat{\mathbf{G}}^T$
given by $\left(\mathbf{0} | \mathbf{B}\right)$ is obtained by
adding to each row of $\tilde{\mathbf{G}}''^T$ a linear combination
of rows from $\tilde{\mathbf{G}}'^T$, in a way that the $m'$
leftmost columns of $\tilde{\mathbf{G}}''^T$ are zeroed-out. It
follows that the statistical properties of $\tilde{\mathbf{G}}''^T$
are inherited by the $m'' \times (k-m')$ submatrix $\mathbf{B}$,
whose elements are hence {uniformly distributed} in $\mathbb
F_q$. {It follows that \eqref{eq:solve}} is solvable if and only if $\mathbf{B}$ is full
rank, i.e., if and only if $\textrm{rank}(\mathbf{B})=k-m'$.}

\subsection{An Efficient Decoding Algorithm}
{We assume next}  the case where the \ac{MDS} code is a
$(n,k)$ \ac{GRS} code {with transposed generator matrix in
Vandermonde form}
\begin{equation}\label{eq:Transpose_generator_RS}
\mathbf{G}'^{T}  = \left( {\begin{array}{*{20}c}
   1 & \beta_1 &  \cdots  & \beta_1^{k-1}  \\
   1 & \beta_2  &  \cdots  & {\beta_2^{k - 1} }  \\
    \vdots  &  \vdots  &  \ddots  &  \vdots   \\
   1 & {\beta_{n}} &  \cdots  & \beta_{n}^{k - 1}  \\
\end{array}} \right),
\end{equation}
where $\beta_i$, $i=1,\ldots ,n$, are $n$ distinct non-{zero} elements
of
 $\mathbb{F}_q$. Efficient decoding can be achieved by {taking advantage of} the structure of
 $\mathbf{G}'$.\footnote{In this
work we consider \ac{MDS} codes based on Vandermonde matrices, but
similar {arguments} hold for \ac{MDS} codes based on Cauchy
matrices.}  {In fact, a Vandermonde matrix
can be inverted} with
quadratic complexity
\cite{Parker64:InverseVandermonde,Turner66:InverseVandermonde,Kaufman:InverseVandermonde,Wertz:InverseVandermonde,Gohberg:FastAlgorithmVandermonde}. {This property has been widely exploited}   for efficient decoding
of \ac{GRS} over erasure channels
\cite{Forney66:concatenatedCodes,mceliece2002theory,brauchle2009systematic,brauchle2011efficient}.
In the following, we first review an efficient method for the
inversion of a Vandermonde matrix based on the LU factorization
\cite{Turner66:InverseVandermonde}. Then, we apply the algorithm of
\cite{Turner66:InverseVandermonde} to the decoding of the proposed concatenated
scheme.

\medskip

\subsubsection{Vandermonde Matrices and Their Inverse}\label{subsec:Vandermonde}

Let us consider a $\gamma \times \gamma$ Vandermonde matrix
\begin{equation}
\mathbf{V}  = \left( {\begin{array}{*{20}c}
   1 & {x_1 } &  \cdots  & { x_1^{\gamma  - 1} }  \\
   1 & {x_2 } &  \cdots  & { x_2^{\gamma  - 1} }  \\
    \vdots  &  \vdots  &  \ddots  &  \vdots   \\
   1 & {x_\gamma  } &  \cdots  & { x_\gamma^{\gamma  - 1} }  \\
\end{array}} \right)\nonumber
\end{equation}
where $x_i$, $i=1,\dots,\gamma$, are $\gamma$
distinct non-zero elements of $\mathbb{F}_q$. In the following, $\gamma$ will be referred to as
the \emph{degree} of the Vandermonde matrix.

The inverse of a $\mathbf V$ matrix can be efficiently computed
according to \cite{Turner66:InverseVandermonde} by means of two
recursions. In particular, the inverse matrix $ \mathbf{V}^{- 1}$
can be obtained as
\[
 \mathbf{V}^{- 1}= \mathbf{U}^{-1} \mathbf{L}^{ - 1}
\]
where
$\mathbf{U}$ is an upper triangular matrix whereas $\mathbf{L}$ is a lower
triangular matrix. The coefficients $l_{i,j}$ of $\mathbf{L}^{-1}$
 are given by
\begin{equation}
l_{i,j}=
\prod\limits_{h = 1,h \ne j}^i {\frac{1}{{x_j  - x_h }}}  \qquad  j \le i,\,\, i>1
\nonumber
\end{equation}
with $l_{1,1}=1$ and $l_{i,j}=0$ for $j>i$. Note that, for the
$j$-th column of $\mathbf{L}^{-1}$, the elements below the
main diagonal can be computed according to the recursion
\[
 l_{i,j}  =
 \frac{l_{i - 1,j}}{{x_j  - x_i }}
 \]
for $i=j+1,\dots,\gamma$, after computing $l_{j,j}$. {
Similarly, the coefficients $u_{i,j}$ of $\mathbf{U}^{-1}$ are given by
\begin{equation}
u_{i,j}=\left\{\begin{array}{lll}
u_{i - 1,j - 1}  - u_{i,j - 1} x_{j - 1} & \qquad & j > i >1\\
- u_{i,j - 1} x_{j - 1}  & \qquad & j > i, i=1
\end{array}\right.
\nonumber
\end{equation}
with $u_{i,i}=1$ and $u_{i,j}=0$ for $j<i$.}
The complexity of
computing $\mathbf{L}^{-1}$ and $\mathbf{U}^{-1}$ is $\mathcal
O(\gamma^2)$.

Let us denote with $ \msr J' =\left\{ { j_{1} ,j_{2} , \dots ,j_{
m'} } \right\} $ any set of $m' \le n$ indexes of rows of
$\mathbf{G}'^{T}$. Consider the square submatrix $\mathbf V$ of
$\mathbf{G}'^{T}$ composed by the $m'$ rows (shortened to their
first $m'$ elements) of $\mathbf{G}'^{T}$ with indexes in $\msr J'$,
\[
\mathbf V=\left( {\begin{array}{*{20}c}
   1 & \beta_{j_{1}} &  \cdots  & \beta_{j_{1}}^{m'-1}  \\
   1 & \beta_{j_{2}}  &  \cdots  & {\beta_{j_{2}}^{m'-1} }  \\
    \vdots  &  \vdots  &  \ddots  &  \vdots   \\
   1 & {\beta_{j_{m'}}} &  \cdots  & \beta_{j_{m'}}^{m'-1}  \\
\end{array}} \right).
\]
Note that $\mathbf V$ is always a  Vandermonde matrix
of degree $m'$, with elements $ x_i^{t - 1}  =
\beta_{j_i}  ^{t - 1}$, for $i,t=1,\dots,m'$. {This observation leads to the following decoding algorithm.}

\medskip

\subsubsection{Decoding Algorithm}\label{subsec:dec_algorithm}

Decoding can be performed with complexity $\mathcal O (k^2)$ {(equivalently, with a $\mathcal O (k)$ cost)} if $m' \ge k$
symbols from the \ac{MDS} code have been received. {In fact, this is the complexity of inverting a Vandermonde matrix of degree $k$.}
If $m'=0$, {the decoding complexity is equivalent to that of \ac{LRFC} decoder, thus cubic in $k$} (resulting in a $\mathcal O (k^2)$ cost), {which is the complexity of applying the \ac{GE} algorithm to solve a linear system of at least $k$ equations in $k$ unknowns.}

\begin{figure}[hb]
\begin{center}
\includegraphics[width=0.6\columnwidth,draft=false]{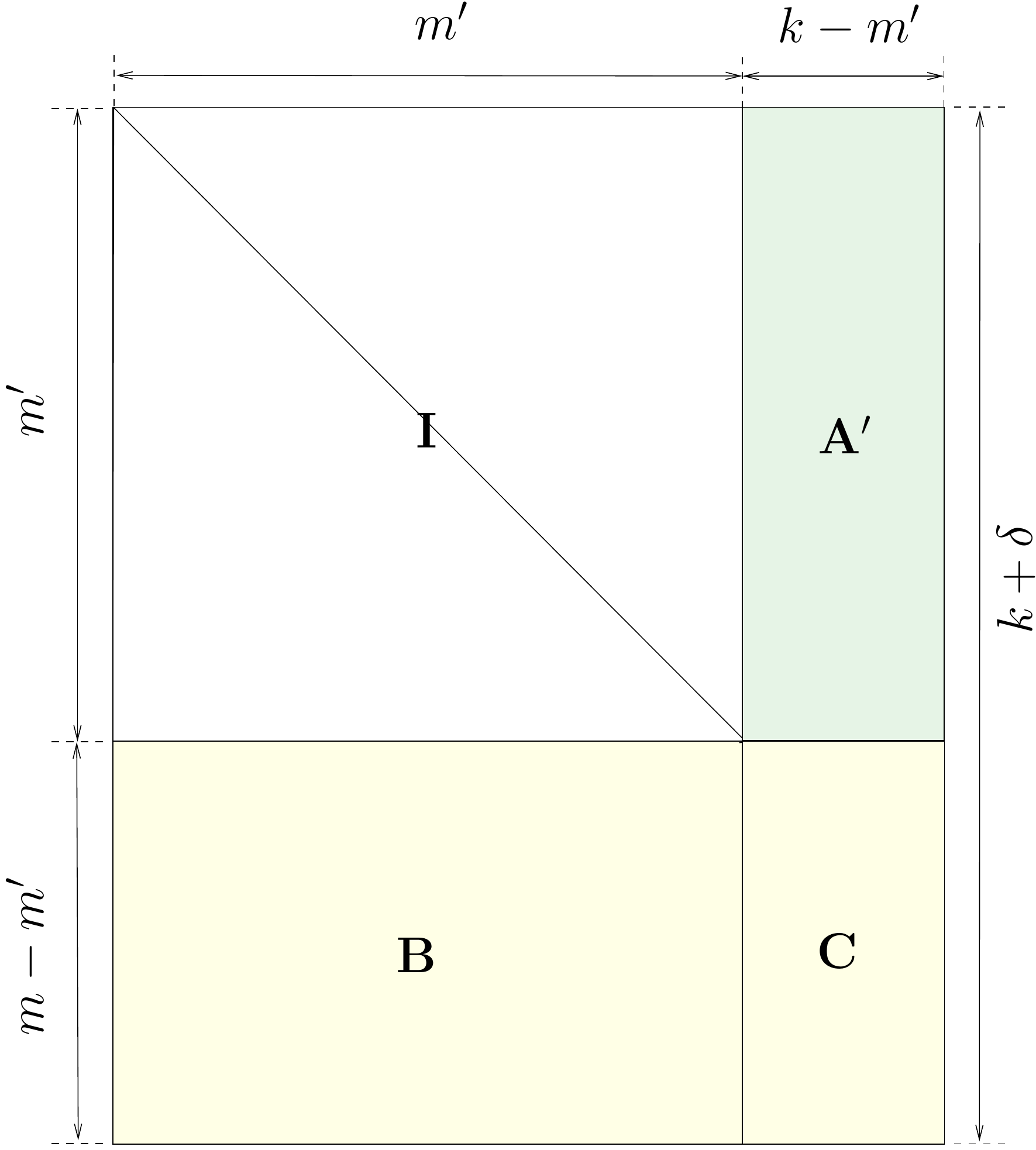}
\centering \caption{Matrix of the system of equations in \eqref{eq:G_partition} after the multiplication with $\mathbf{M}$.}\label{fig:subfig1}
\end{center}
\end{figure}

Let us consider the case {where $0 < m' < k$ symbols of the
\ac{MDS} code have been collected, among the $m \ge k$ received symbols.}
{We can define $m'$ as a fraction of $k$,  $m' = \xi k$, with $0 < \xi < 1$.}  The matrix
$\tilde{\mathbf{G}}^T$ can be written as  \[ \tilde{\mathbf{G}}^T=\left(\begin{array}{ccc}
\mathbf{V} & \vline & \mathbf{A} \\\hline
\mathbf{B} & \vline & \mathbf{C}\\
\end{array}\right)\] where $\mathbf{V}$ is a Vandermonde matrix of degree $m'$, whereas $\mathbf{A}$, $\mathbf{B}$, $\mathbf{C}$ have respective sizes $m' \times (k-m')$, $(m-m') \times m'$, $(m-m') \times (k-m')$. An efficient decoding algorithm can be derived by inverting $\mathbf{V}$  according to the algorithm presented in Section \ref{subsec:Vandermonde}. {Given the matrix $\mathbf{V}^{-1}$, $\tilde{\mathbf{G}}^T$ can be multiplied by a full-rank matrix $\mathbf{M}$,  with
 \[ \mathbf{M} = \left(\begin{array}{ccc} \mathbf{V}^{-1} & \vline & \mathbf{0} \\\hline
      \mathbf{0} & \vline & \mathbf{I}\\
      \end{array}\right),\]
$\mathbf{I}$ being a $(m-m')\times(m-m')$ identity matrix, leading to the matrix depicted in Fig.~\ref{fig:subfig1}.
Accordingly, \eqref{eq:solve} is modified as
\[
\mathbf{M} \cdot \tilde{\mathbf{G}}^T  \cdot \mathbf{u}^T  =
\mathbf{M} \cdot \mathbf{y}^T.
\]}
The complexity of
multiplying {the $m' \times m'$ matrix} $\mathbf{V}^{-1}$ with the {matrix} $\mathbf{A}$, leading to
the $m' \times (k-m')$ matrix $\mathbf{A}'$, is
{$\mathcal O({m'}^2(k-m'))$, which is the complexity of performing standard matrix multiplications}.

\begin{figure}[ht]
\begin{center}
\includegraphics[width=0.6\columnwidth,draft=false]{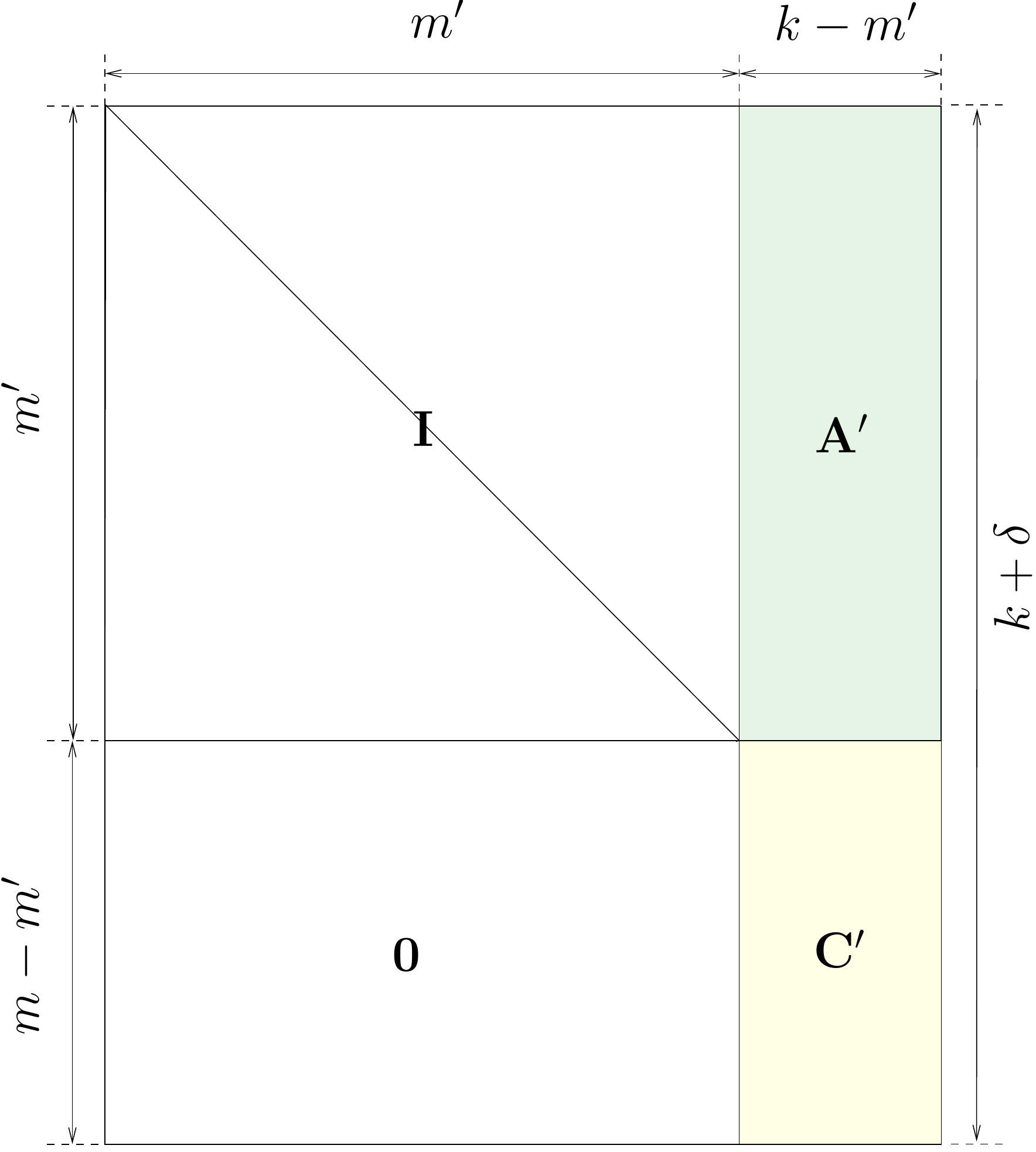}
\centering \caption{Matrix of the system of equations in \eqref{eq:G_partition} with $\mathbf{B}=\mathbf{0}$.}\label{fig:subfig2}
\end{center}
\end{figure}

Referring to Fig.~\ref{fig:subfig1}, the $i$-th row of the matrix $\mathbf{B}$ (for $i=1,\dots,m-m'$) can be zeroed-out by adding to it a linear combination of the $m'$ rows of $\left(\mathbf{I} | \mathbf{A}'\right)$.
The complexity of zeroing-out $\mathbf{B}$ is {$\mathcal O((m-m')m'(k-m'))$}, and the resulting system matrix is depicted in Fig.~\ref{fig:subfig2}. {In fact, $\mathbf{B}$ is a random matrix with entries uniformly distributed in $\mathbb F_q$. Due to the linear combinations performed to zero-out the matrix $\mathbf B$, the matrix $\mathbf{C}$ results in
in a new matrix $\mathbf{C}'$. Thus, a \ac{GE} step is performed on the matrix $\mathbf{C}'$ in order to recover the $k-m'$ symbols involved in the lower part of the system of equations with complexity $\mathcal O((k-m')^3)$}. {Finally, back-substitution is applied in order to recover the $m'$ symbols involved in the upper part of the system of equations with complexity $\mathcal O(m'(k-m'))$.}

{Since $m'$ is a fraction of $k$, the complexity of the proposed algorithm is $\mathcal O(k^3)$ (i.e., $\mathcal O(k^2)$ cost). However, the constant hidden by the $\mathcal O$-notation becomes smaller as $m'$ approaches $k$ (in the limit case where $m'=k$, the decoding complexity is actually quadratic in $k$).}

\section{Performance Analysis} \label{sec:bounds} Based on the bounds
{\eqref{eq:tightbounds}}, tight upper and lower bounds for
the decoding failure probability of the fountain coding scheme can
be derived in case of a memory-less erasure channel. The decoding failure
probability $P_f=\Pr\{F\}$, where $F$ denotes the decoding
failure event is defined as the probability that the source block
$\mathbf{u}$ cannot be recovered out of a set of received symbols.
{We focus on the case where the linear block code
used in concatenation with the \ac{LRFC} is maximum distance
separable (MDS). When binary codes will be used, we assume
$(k+1,k)$ \ac{SPC} codes. When operating
on higher order finite fields, we consider  \ac{GRS}
codes.}

\begin{figure}[h]
\begin{center}
\includegraphics[width=0.95\columnwidth,draft=false]{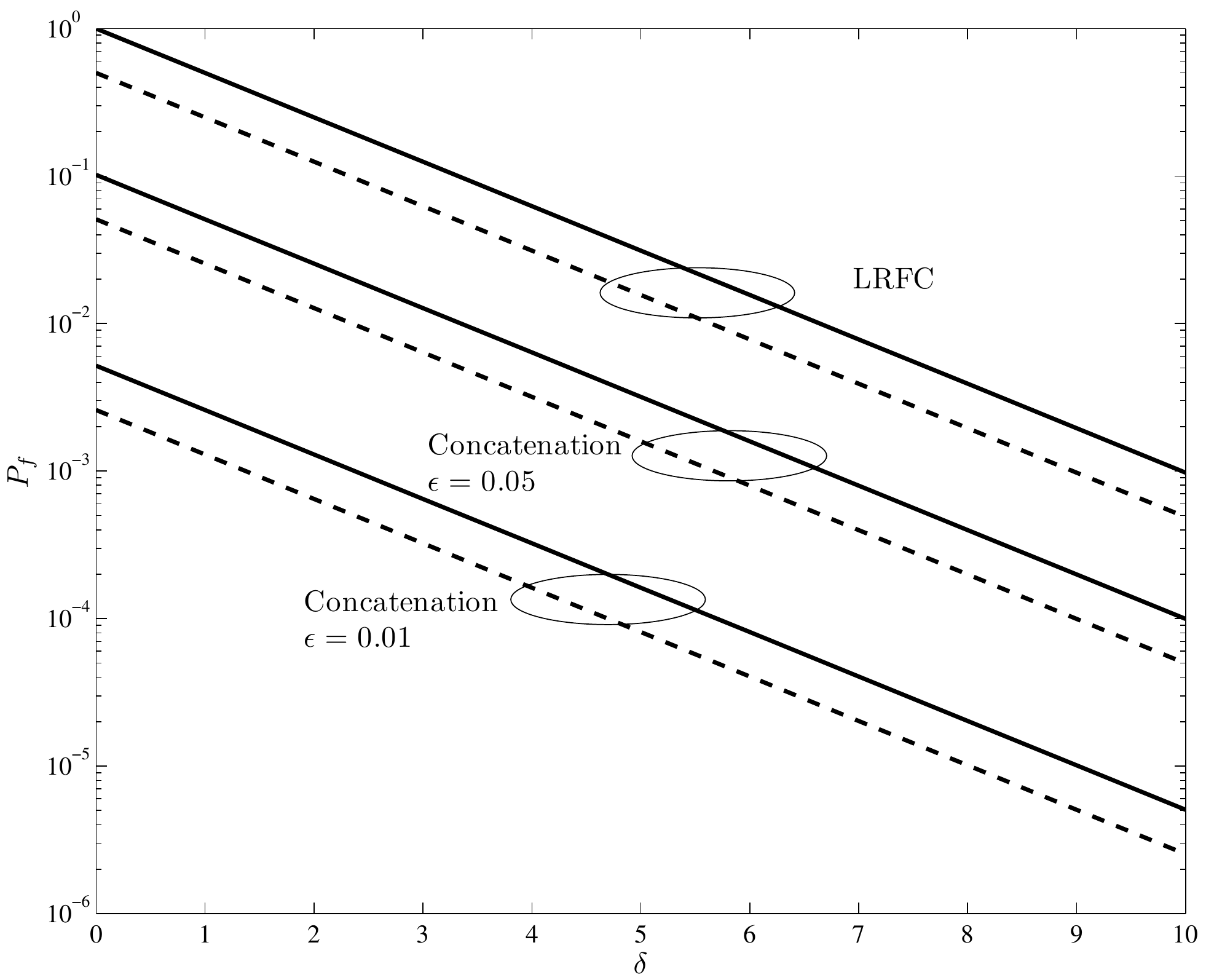}
\centering \caption{$P_f(\delta,\epsilon)$ vs. overhead  for a concatenated code built using a $(11,10)$ \ac{SPC} code over $\mathbb {F}_{2}$ for  different values of $\epsilon$.
Upper bounds are represented by solid lines and lower bounds are represented by dashed lines.} \label{GF_2}
\end{center}
\end{figure}

\begin{figure}[h]
\begin{center}
\includegraphics[width=0.95\columnwidth,draft=false]{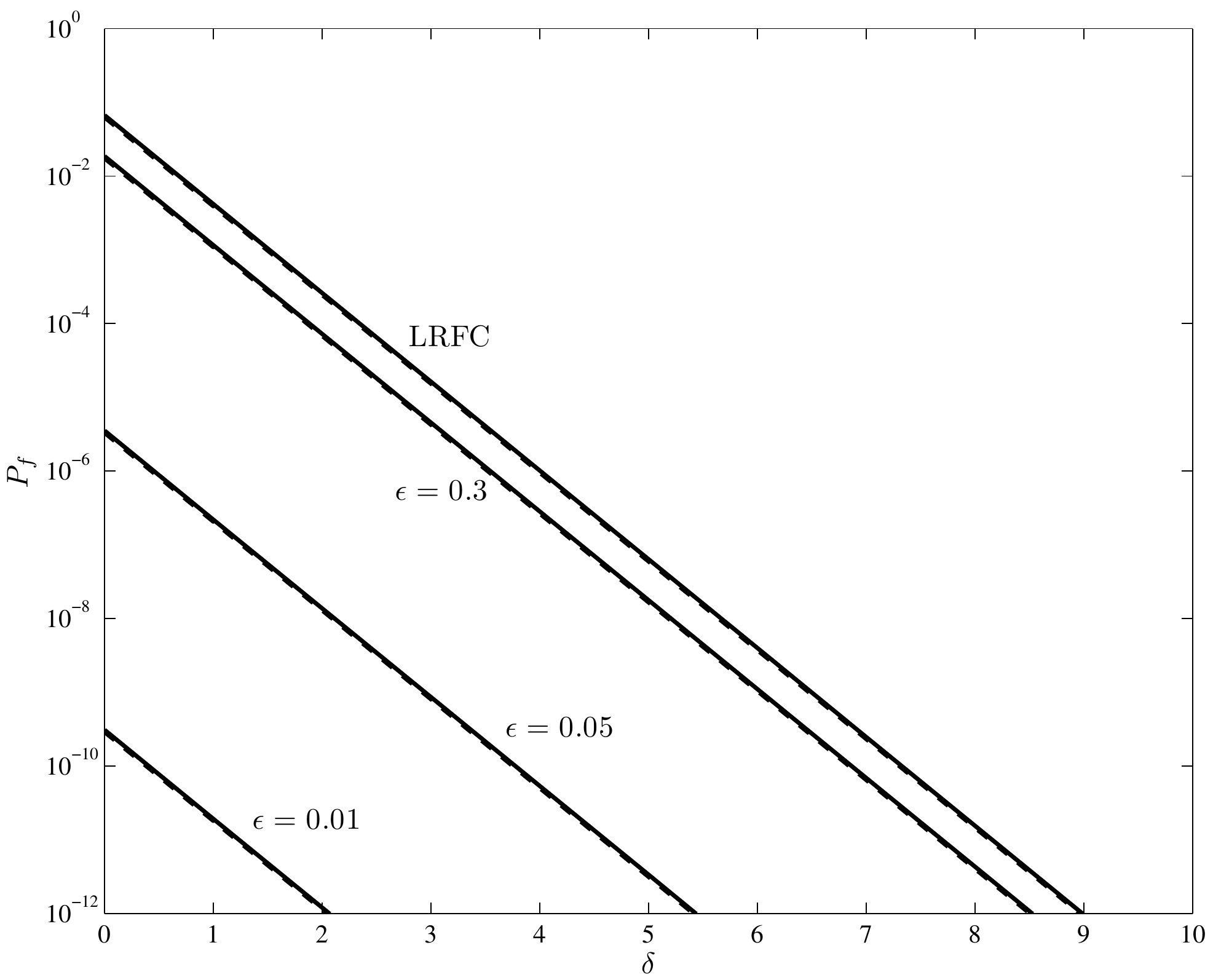}
\centering \caption{$P_f(\delta,\epsilon)$ vs.  overhead  for a concatenated code built using a $(15,10)$ \ac{RS} over $\mathbb {F}_{16}$ for  different values of $\epsilon$.
Upper bounds are represented by solid lines and lower bounds are represented by dashed lines.} \label{GF_16}
\end{center}
\end{figure}

Suppose now that an encoded sequence $\mathbf{c}$ composed of $l \ge n$ symbols
is transmitted over an erasure channel with erasure
probability of $\epsilon$.\footnote{The case $l < n$ is not considered since it
is equivalent to shortening the linear block code.} The probability that at least $k$ symbols
out of the $n$ symbols produced by the linear block code encoder are
received is given by
\begin{equation}
Q(\epsilon)=\sum_{i=k}^n {n \choose i} (1-\epsilon)^i
\epsilon^{n-i}.
\nonumber
\end{equation}
Hence, with a probability $P(\epsilon)=1-Q(\epsilon)$ the
receiver would need to collect symbols encoded by the \ac{LRFC}
encoder to recover the source block. Assuming that the receiver collects
$m=k+\delta$ symbols, out of which only $m'<k$ have been produced by
the linear block encoder, the conditional decoding failure
probability can be expressed as
\begin{equation}
\Pr\{F|m',m'<k,\delta\}=\Pr\{\textrm{rank}(\mathbf{B})<k-m'\}.\label{eq:cond_F_prob_1}
\end{equation}
{Note that $\mathbf{B}$ is a $m'' \times (k-m') = (k+\delta-m')
\times (k-m')$ random matrix having $\delta$
rows in excess w.r.t. the number of columns. We can thus
replace  \eqref{eq:cond_F_prob_1} in \eqref{eq:tightbounds}, obtaining
the bounds
\begin{equation}
q^{-\delta-1}\leq\Pr\{F|m',m'<k,\delta\}<\frac{1}{q-1}q^{-\delta}.\label{eq:cond_F_prob_2_bounds}
\end{equation}
{Observing that the the bounds in  \eqref{eq:tightbounds} are independent from the size of the matrix (i.e., they
depend only on the overhead), the conditioning
on $m'$ can be removed from \eqref{eq:cond_F_prob_2_bounds}, leaving}
\[
q^{-\delta-1}\leq\textrm{Pr}\{F|m'<k,\delta\}<\frac{1}{q-1}q^{-\delta}.
\]
The failure probability can be written as a function of $\delta$ and $\epsilon$ as
\begin{equation}
\begin{array}{cc}
P_f(\delta,\epsilon)=& \Pr\{F|m'<k,\delta\}\Pr\{m'<k\}\\
&+\Pr\{F|m'\geq k,\delta\}\Pr\{m'\geq k\}
\label{eq:general_bound}
\end{array}
\end{equation}
where $\Pr\{F|m'\geq k,\delta\}=0$ (since at least $k$ symbols output by the \ac{MDS} code encoder have been collected)} and
$\Pr\{m'<k\}=P(\epsilon)$. It results that
\begin{equation}
P(\epsilon) q^{-\delta-1}\leq
P_f(\delta,\epsilon)<P(\epsilon)\frac{1}{q-1}q^{-\delta}.\label{eq:final_bounds}
\end{equation}
From an inspection of \eqref{eq:tightbounds} and
{\eqref{eq:final_bounds}}, one can note how the bounds on the failure
probability of the concatenated scheme are scaled down by a factor
$P(\epsilon)$, which is a monotonically increasing
function of $\epsilon$. It follows that, when the channel conditions
are \emph{bad} (i.e., large $\epsilon$) $P(\epsilon)\rightarrow
1$, and the bounds in {\eqref{eq:final_bounds}} tend to coincide with
the bounds in \eqref{eq:tightbounds}. When the channel conditions
are \emph{good} (i.e., small $\epsilon$), most of the time $m'\geq
k$ symbols produced by the linear block encoder are received,
leading to a decoding success (recall the assumption of \ac{MDS}
code). In these conditions, $P(\epsilon)\ll 1$, and according to
the bounds in  {\eqref{eq:final_bounds}} the failure probability may
decrease by  several orders of magnitude. {Since the probability of
decoding failure of the concatenated scheme is a function of the erasure probability, the scheme is
not universal anymore. More specifically, at low channel erasure probabilities the proposed scheme will outperform universal (random) \acp{LRFC}, whereas for large erasure probabilities it will perform as a universal \ac{LRFC}.}
Fig. \ref{GF_2} shows the probability of decoding failure as a
function of the number of overhead symbols for a concatenated code
built using a $(11,10)$ \ac{SPC} code {over} $\mathbb {F}_2$. It can be
observed how, for lower erasure probabilities, {the gain in performance of the concatenated code with respect to a \ac{LRFC}
 increases}. For
$\epsilon=0.01$ the decoding failure probability is more than $2$
orders of magnitude lower {than that of a \ac{LRFC}}. Fig. \ref{GF_16} shows the probability of
decoding failure vs. the number of overhead symbols for the
concatenation of a $(15,10)$ \ac{RS} and a \ac{LRFC} over $\mathbb
{F}_{16}$. The performance of the concatenated code is compared with
that of the \ac{LRFC} built on the same field for different erasure
probabilities. In this case the decrease in terms of probability of
decoding failure is {even more evident than the one of the binary case}. For a channel with an erasure probability
$\epsilon=0.05$, the probability of decoding failure of the
concatenated scheme is $4$ orders of magnitude lower than {that of} the
\ac{LRFC}.

{The analysis provided in this section is also
valid if the \ac{LRFC} is replaced by a Raptor
code.\footnote{{As observed in \cite{Liva10:fountain}, short Raptor
codes over $\mathbb {F}_{q}$ show performance close to those of
\acp{LRFC} constructed over the same field, down to moderate-low
error rates. We therefore expect that the results attained by the
proposed concatenation could be closely approached by replacing the
non-binary \ac{LRFC} with a non-binary Raptor code.}} In order to
calculate the performance of such a concatenated code one has to
{replace} in \eqref{eq:general_bound} the term
$\Pr\{F|m'<k,\delta\}$ by the probability of decoding failure
of the Raptor code. {Also in this case, the failure probability of the
concatenated scheme is reduced by a factor $P(\epsilon)$ with respect to that of the Raptor code}.}

\begin{figure}[ht]
\begin{center}
\includegraphics[width=0.95\columnwidth,draft=false]{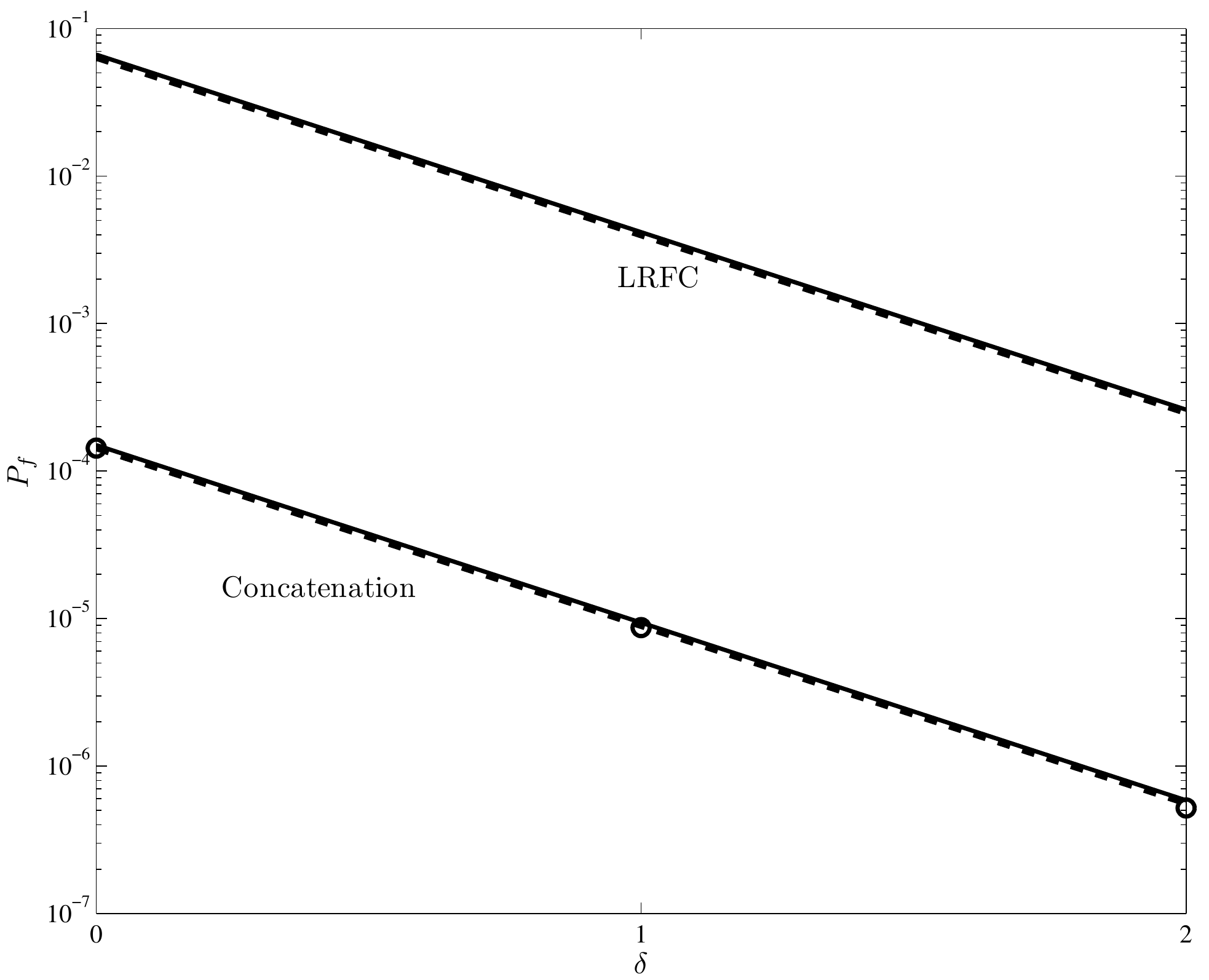}
\centering \caption{$P_f(\delta,\epsilon)$ vs. overhead  for a
the concatenation of a $(15,10)$ \ac{RS} and \ac{LRFC} over $\mathbb {F}_{16}$ and $\epsilon=0.1$. Upper and lower bounds are represented by solid and dashed lines, respectively. The markers '$\circ$' denote simulations.} \label{GF_16_sim}
\end{center}
\end{figure}

\begin{figure}[ht]
\begin{center}
\includegraphics[width=0.95\columnwidth,draft=false]{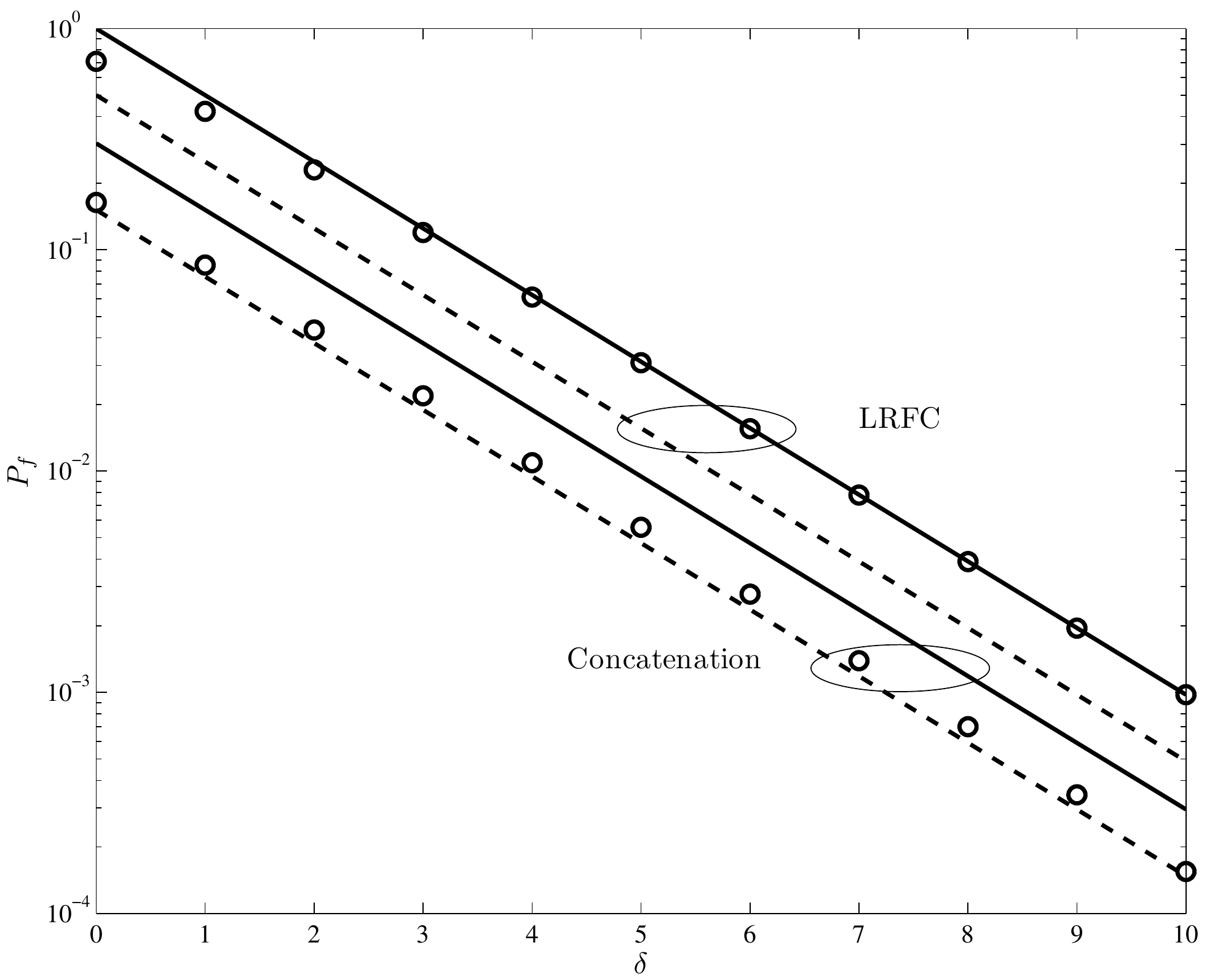}
\centering \caption{$P_f(\delta,\epsilon)$ vs. overhead symbols for a
the concatenation of a $(11,10)$ \ac{SPC} code and a \ac{LRFC} over $\mathbb {F}_{2}$ and $\epsilon=0.1$. Upper bounds are represented by solid lines and lower bounds are represented by dashed lines. The points marked with '$\circ$' denote actual simulations.} \label{GF_2_sim}
\end{center}
\end{figure}

\section{Numerical Results}\label{sec:results}
{Fig. \ref{GF_16_sim} shows the probability of decoding failure $P_f$, as a function of the overhead $\delta$, obtained via Monte Carlo  simulations. } {The results refer to a concatenation of a
$(15,10)$ \ac{RS} code with a
\ac{LRFC} over $\mathbb {F}_{16}$, for a channel erasure
probability $\epsilon=0.1$. The results are compared with}
the bounds of {\eqref{eq:final_bounds}}. {As expected, the
simulation results tightly match the bounds. Fig. \ref{GF_2_sim}
shows the simulation results for a concatenated code using a
$(11,10)$ parity check code over $\mathbb {F}_{2}$, and a channel with
an erasure probability $\epsilon=0.1$. Also in this case, the results are remarkably close to the bounds.}

\begin{figure}[t]
\begin{center}
\includegraphics[width=0.95\columnwidth,draft=false]{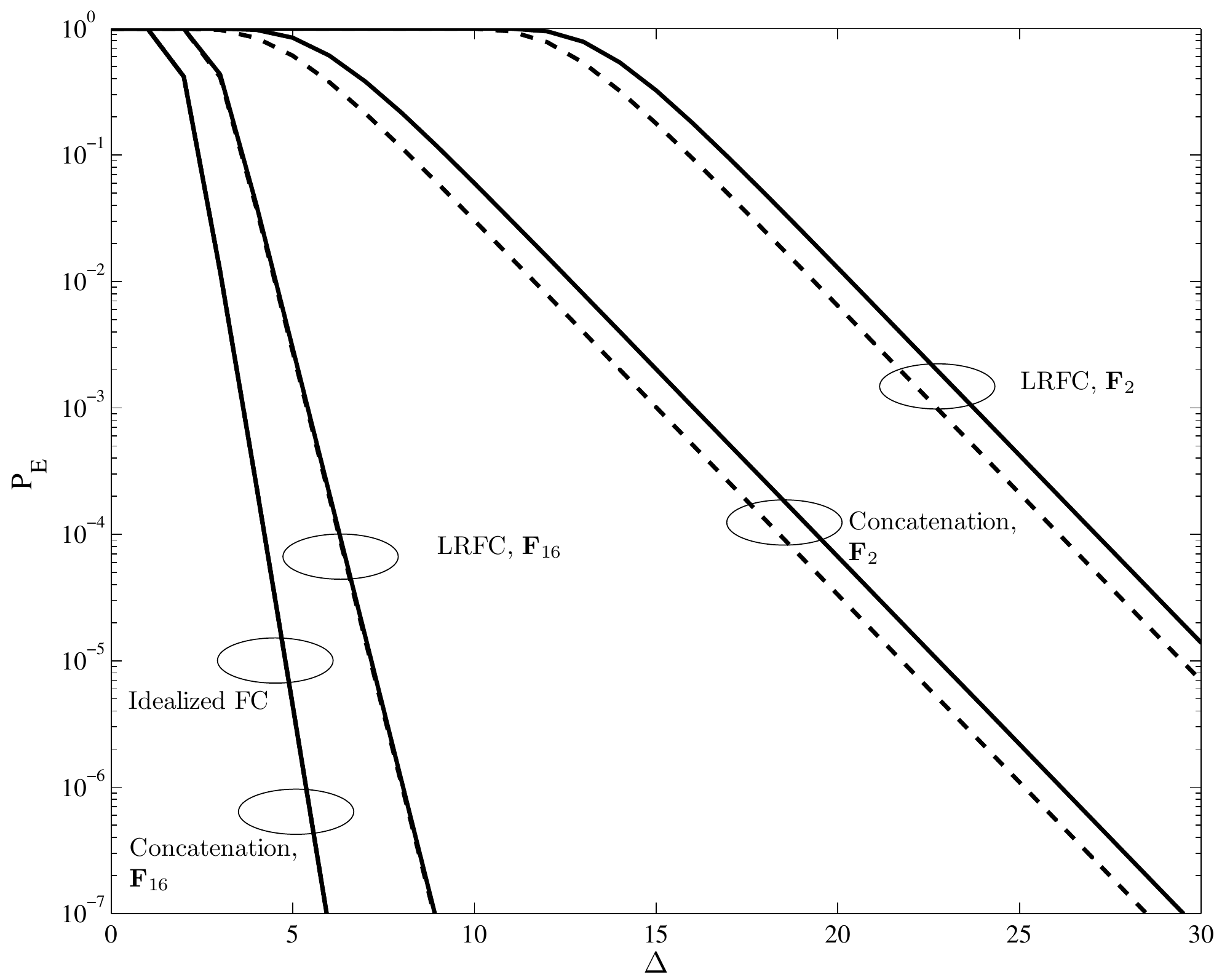}
\centering \caption{ $P_E$ vs. overhead at the transmitter in a system with $N=10000$ receivers and $\epsilon=0.01$. Results are shown for different fountain codes:
\ac{LRFC} in $\mathbb {F}_{2}$, \ac{LRFC} in $\mathbb {F}_{16}$, concatenation of a (11,10) \ac{SPC} code with a \ac{LRFC} code in $\mathbb {F}_{2}$,
and a concatenation of a {$(15,10)$} \ac{RS} code and a \ac{LRFC} code over $\mathbb {F}_{16}$.}\label{sim_sender_side}
\end{center}
\end{figure}

{The performance of the concatenated scheme in a system
with a large  receivers population has been performed. The number of receivers is denoted by $N$.
We considered the erasure channels from the transmitter to the different
receivers to be independent, albeit with an identical erasure probability
$\epsilon$. Furthermore, we assumed that the receivers send an
acknowledgement to the transmitter whenever they successfully
decode the source block. Ideal (error- and delay-free) feedback channels have been
considered. After retrieving all the acknowledgments, the
transmitter stops encoding additional symbols from the source block.}
{We denote next by $\Delta$ the number of symbols transmitted by the sender, in excess with respect to $k$. We refer to $\Delta$ as the transmission overhead. When  $k+\Delta$
symbols have been transmitted, the probability that a specific
receiver gathers exactly $m$ symbols is}
\begin{equation}
\ S\left(\Delta,m\right) = \binom{k+\Delta}{m}(1-\epsilon)^{m}\epsilon^{k+\Delta-m}.
\label{system_prob m}
\end{equation}
The probability of decoding failure at the receiver given that the
transmitter has sent $k+\Delta$ symbols is hence
\begin{align*}
\ P_{e} =& \sum_{m=0}^{k-1}\ S\left(\Delta,m\right)+\\
& + \sum_{m=k}^{k+\Delta}\ S\left(\Delta,m\right) P_{f}(\delta=m-k,\epsilon).
\end{align*}
The probability that at least one receiver is not able to  decode the source block
is thus
\begin{equation}
\ P_E(N,\Delta,\epsilon) = 1-(1-P_{e})^{N}
\label{system_failure_one user}
\end{equation}
{Observe that $P_E(N,\Delta,\epsilon)$ can be easily bounded by means of}  {\eqref{eq:final_bounds}}. {Following this approach, we compare the performance of the proposed concatenation to that of \acp{LRFC} and to that of idealized fountain codes. We assume a system
with $N=10^{4}$ receivers and a channel with an erasure probability
$\epsilon=0.01$. The performance of \ac{LRFC} codes over $\mathbb
{F}_{2}$ and $\mathbb {F}_{16}$ is depicted in Fig. \ref{sim_sender_side} together with that of two
concatenated schemes: A concatenation of a $(11,10)$ \ac{SPC} code
with a \ac{LRFC} code over $\mathbb {F}_{2}$,} and a concatenation of a
$(15,10)$ \ac{RS} code and a \ac{LRFC} code over $\mathbb {F}_{16}$.
{It can be seen how the concatenated scheme in $\mathbb {F}_{2}$
outperforms the binary \ac{LRFC}. To achieve $P_E = 10^{-4}$ the concatenated scheme
 needs only $\Delta=20$ overhead symbols whereas the
\ac{LRFC} requires a transmission overhead $\Delta=27$. In the case of
a field order $16$, the concatenated
code shows a performance very close to that of an idealized
fountain code.}

\section{Conclusions}\label{sec:conc}
A novel coding scheme has been introduced. The scheme
consists of a parallel concatenation of a \ac{MDS} block code with a
\ac{LRFC} code, both constructed over the same field. The performance of the concatenated coding scheme
has been analyzed through derivation of tight bounds on the
probability of decoding failure as a function of the receiver overhead. It
has been shown how under \ac{ML} decoding the concatenated scheme performs as well as
\ac{LRFC} codes in channels characterized by high erasure
probabilities, whereas {it provides} failure probabilities {lower than those of \ac{LRFC} codes}  by
several orders of magnitude at moderate/low erasure probabilities. An efficient decoding algorithm has been introduced for the case in which the generator
matrix of the \ac{MDS} block code is in Vandermonde form. Finally, the complexity of the proposed decoding algorithm has been analyzed, {showing remarkable complexity savings at moderate/low erasure probability regimes.}


\appendix[Performance in the Finite Rate Setting]\label{sec:appendix}

Fountain codes are often used in a finite rate setting as conventional erasure correcting codes \cite{MBMS05:raptor}, \cite{DVB-SH:raptor}. In this context, the main advantage in the use of fountain codes with respect to block erasure correcting codes stems from the possibility of adapting code rate and block length to the transmission needs (e.g., channel conditions) in a flexible manner. In the following, we derive tight upper bounds on the block error probability for the codes proposed in the paper, for the general case where the block code {$\mathcal C'$ is not \ac{MDS}.}

In order to characterize the block error probability of a code under \ac{ML} decoding  we first seek for the \ac{WEF} of the code.  The coding scheme proposed in this work is a parallel concatenation of a linear block code and a \ac{LRFC}, which for a finite rate setting is a random generator matrix code.  Let us denote as $\msr{C}(\mathcal{C}',k,l,q)$ the ensemble of codes obtained by a parallel-concatenation of a $(n,k)$ linear block code over $\mathbb{F}_q$, $\mathcal{C}'$, with all possible realizations of a \ac{LRFC}, where $k$ is the number of source symbols, $l$ is the total number of output symbols and $q$ is the finite field order. The rate for the codes in the ensemble is, therefore, $r=k/l$. We denote as $\mathcal A_i(X)$ the \ac{C-OWEF} averaged over the ensemble $\msr{C}(\mathcal{C}',k,l,q)$  conditioned to the input source block having
weight $i$,
\begin{equation}
\mathcal A_i(X)= \sum_{w=1}^{l} \mathcal A_{i,w}X^w
\nonumber
\end{equation}
where $\mathcal A_{i,w}$ is the average number of codewords of Hamming weight $w$ produced by Hamming weight-$i$ inputs. For the ensemble of parallel-concatenated codes the average \ac{C-OWEF}  can be written as
\begin{equation}
\mathcal A_i(X)= \frac{A_i^{\mathcal{C}'}(X) \mathcal A_i^{\msr {L}(k,h,q)}(X)  }{{k \choose i}},
\label{eq:iowe}\nonumber
\end{equation}
where $A_i^{\mathcal{C}'}(X)$ is the \ac{C-OWEF} of the linear block code, and $\mathcal A_i^{\msr {L}(k,h,q)}(X)$ is the average \ac{C-OWEF} of the ensemble $\msr {L}(k,h,q)$, being $\msr {L}(k,h,q)$ the ensemble of linear block codes over $\mathbb{F}_q$ with $k \times h$ generator matrix $\mathbf{G}''$, with $h=l-n$. Assuming $A_i^{\mathcal{C}'}(X)$ known{\footnote{{In general, the derivation of the \ac{C-OWEF} $A_i^{\mathcal{C}'}(X)$ for a code is not trivial, unless the code $\mathcal{C}'$ (or its dual code) has small dimension \cite{MacWillimas77:Book}.}}}, the derivation of  $\mathcal A_{i,w}$ reduces to the calculation of  $\mathcal A_i^{\msr {L}(k,h,q)}(X)$.

We denote by $\mathcal A_{i,w}^{\msr {L}(k,h,q)}$  the average number of codewords of Hamming weight $w$ produced by Hamming weight-$i$ inputs for the ensemble $\msr {L}(k,h,q)$ which is given by:
\begin{equation}
\mathcal A_{i,w}^{\msr {L}(k,h,q)} = {{k \choose i}} {{h \choose w}} p_i^w \left( 1-p_i \right) ^{h-w},
\label{eq:iowe_LRFC_0}\nonumber
\end{equation}
where $p_i(q)$ the probability for each of the $h$ output symbols having a non-zero value conditioned to having an input of Hamming weight $i$. Assuming the coefficients of $\mathbf{G}''$ are picked with uniform probability over $\mathbb{F}_q$, we have that{\footnote{{Note that when $i=0$ the encoder input is given by the all-zero word. Thus, the encoder output is zero with probability $1$ due to the linearity of the code ensemble $\msr {L}(k,h,q)$.}}}
\begin{equation}
\begin {array}{lll}
&p_i = \frac{q-1}{q}&, \; i\neq 0 \\
&p_i = 0&, \; i = 0.
\end{array}
\label{eq:iowe_LRFC_1}\nonumber
\end{equation}

Finally, from the average \ac{C-OWEF}, $\mathcal A_i(X)$, the average \ac{WEF} $\mathcal A(X)$  can be computed as
\[
\mathcal A(X) = \sum_w \mathcal A_w X^w
\]
being $A_w$ the average number of codewords of Hamming weight $w$, $\mathcal A_w = \sum_i \mathcal A_{i,w}$.

The average \ac{WEF} of the concatenated ensemble can be used now to derive tight upper bounds on the expected block error probability for the codes of the ensemble. Let $\mathcal{C}$ be a
 linear block code  belonging to the ensemble $\msr{C}(\mathcal{C}',k,l,q)$. The block error probability averaged over the ensemble can be upper bounded as~\cite{CDi2001:Finite,Liva2013}
\begin{align}\label{eq:bound_Gavg}
&\Bbb{E}_{\msr{C}(\mathcal{C}',k,l,q)} \left[P_B(\mathcal{C},\epsilon)\right]\leq
P^{(\mathsf S)}_{B}(l,k,\epsilon) \nonumber \\ & + \sum_{e=1}^{l-k} {l \choose e} \epsilon^e (1-\epsilon)^{l-e} \min \left\{1, \sum_{w=1}^e {e \choose w} \frac{\mathcal A _w}{{l \choose w}}\right\}
\end{align}
where $P^{(\mathsf S)}_{B}(l,k,\epsilon)$ is the Singleton bound
\begin{equation}\label{eq:bound_S}
P^{(\mathsf S)}_{B}(l,k,\epsilon)= \sum_{e=l-k+1}^l {l \choose e} \epsilon^e (1-\epsilon)^{l-e}.
\end{equation}

\begin{figure}[h]
\begin{center}
\includegraphics[width=0.95\columnwidth,draft=false]{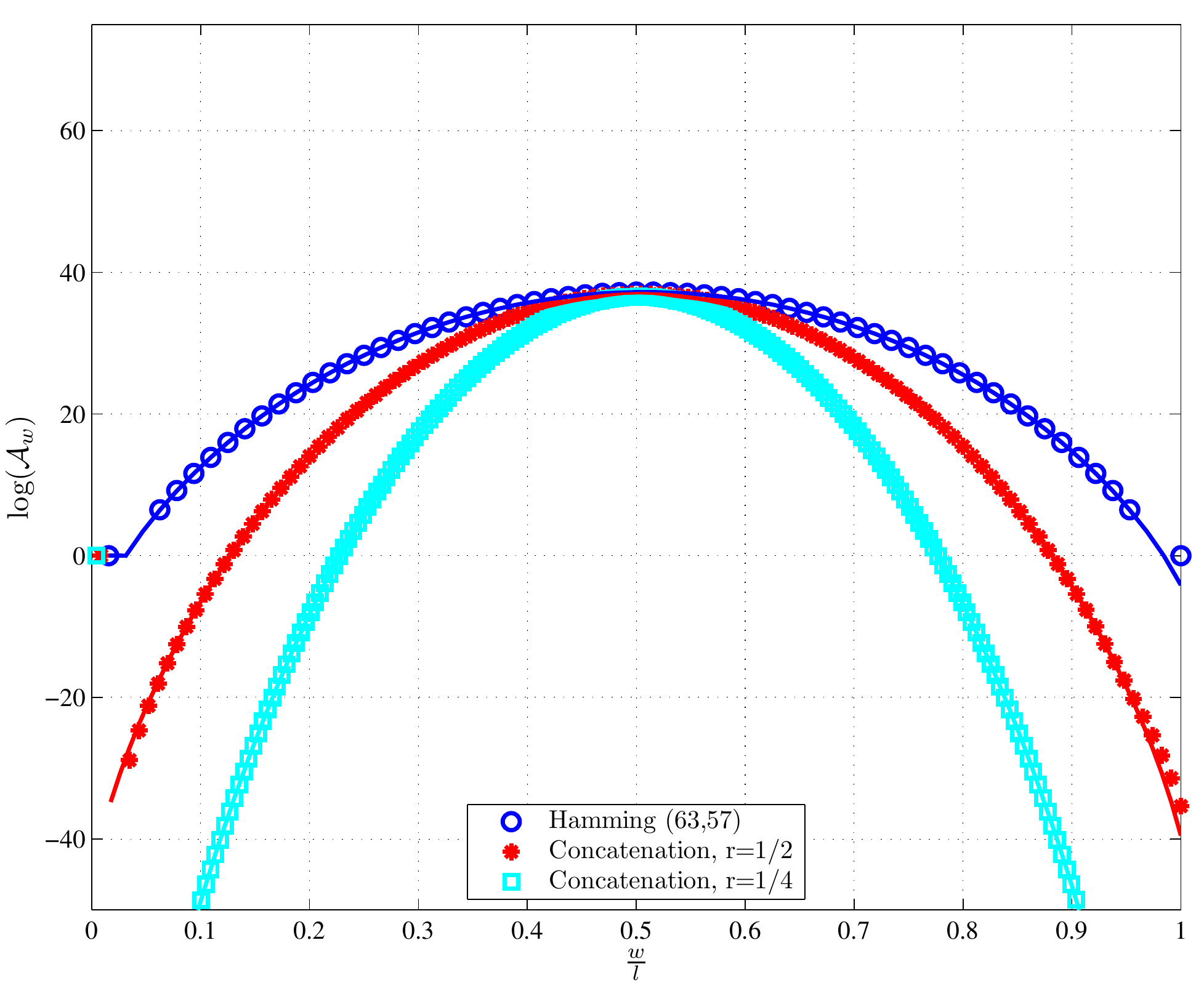}
\centering \caption{ $\log (\mathcal A_w)$ vs. $\frac{w}{l}$ for the concatenation of a (63,57) Hamming code with a \ac{LRFC} code in $\mathbb {F}_{2}$. The round markers represent the distance spectrum for the Hamming code. The asterisks and squares represent the distance spectrum of the concatenated scheme with rates $r=\frac{1}{2}$ and $r=\frac{1}{4}$ respectively. The solid lines represent the average distance spectrum for a random generator matrix code (equivalent to a \ac{LRFC} in a finite rate setting). }\label{dist_spectrum}
\end{center}
\end{figure}

\begin{figure}[h]
\begin{center}
\includegraphics[width=0.95\columnwidth,draft=false]{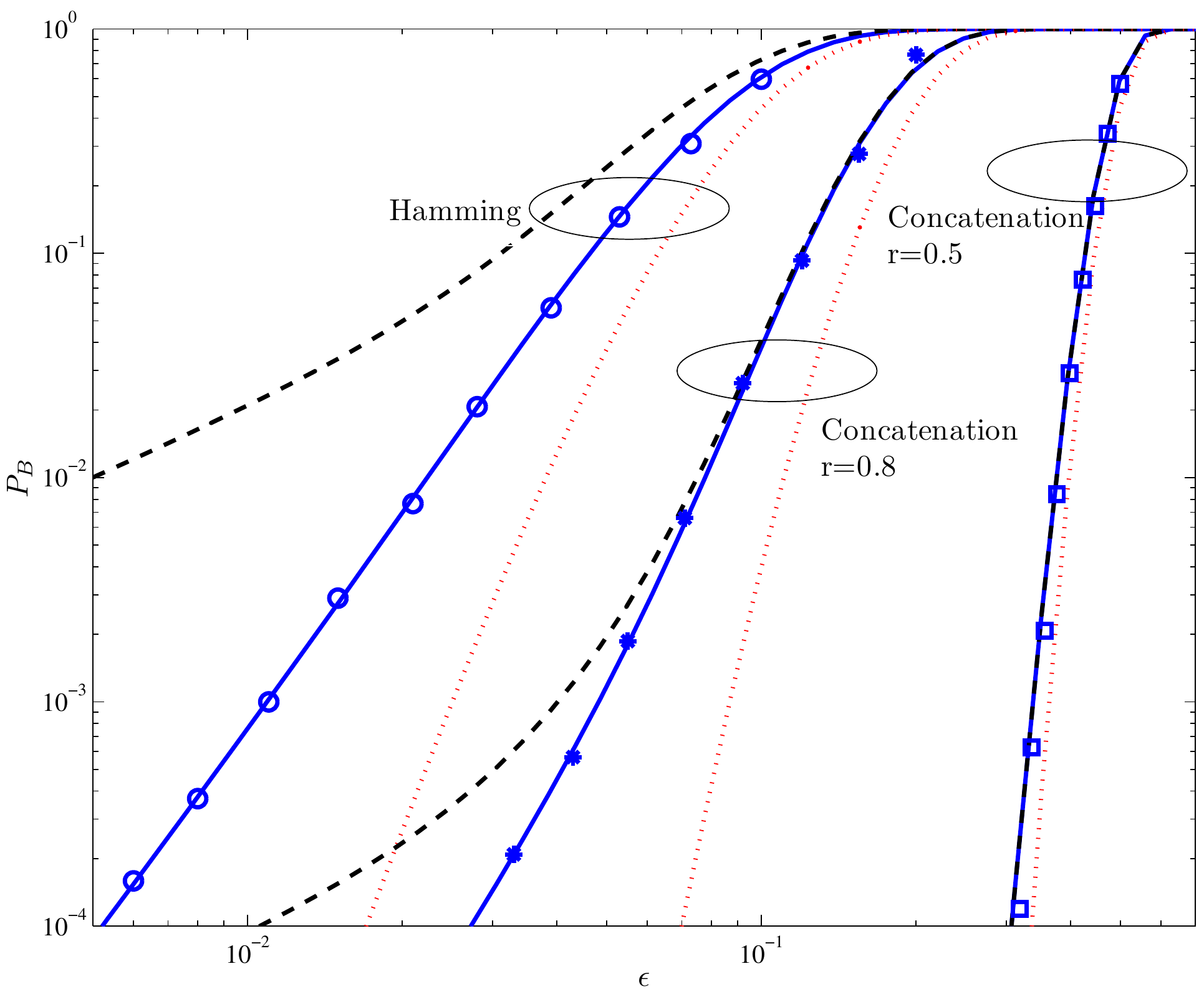}
\centering \caption{ $P_B$ vs. erasure probability $\epsilon$ for the concatenation of a (63,57) Hamming code with a \ac{LRFC} code in $\mathbb {F}_{2}$. The markers represent the result of Monte Carlo simulations. The solid line represents the upper bound in \cite{CDi2001:Finite}, and the black dashed and red dotted lines represent the Berlekamp random coding bound and the Singleton bound respectively.}\label{sim_hamming}
\end{center}
\end{figure}

As an example, consider the concatenation where the block code is a binary $(63,57)$ Hamming code. {Recall that the \ac{C-OWEF} $\mathcal A_i(X)$ of a $(n=2^t-1,k=n-t)$ Hamming code \cite{Chiaraluce:hamming} can be derived from
\begin{align*}
\mathcal A(x,X) = &\frac {(1+x)^{2^{t-1}-t-1}} {2^t} \times  \Big( 2^t(1-x)^{2^{t-1}-t} (1-xX)^t \\
& - (1-x)^{2^{t-1}} (1+X)^t + (1+x)^{2^{t-1}} (1+X)^t \Big)
\end{align*}
where $\mathcal A(x,X)=\sum_i \mathcal A_i(X)x^i$.}
Fig. \ref{dist_spectrum} shows the average distance spectrum of the concatenated code. The markers represent the distance spectrum of the concatenated code, whereas the solid lines represent the average distance spectrum for the ensemble of \ac{LRFC} with rate equal to the concatenated scheme. Fig. \ref{sim_hamming} shows the upper bounds on the expected block error probability of the ensemble, $P_B$, as a function of the channel erasure probability $\epsilon$ for different coding rates. The solid lines represent the upper bound on the block error probability in \eqref{eq:bound_Gavg}, and the dashed black and dotted red lines represent respectively the Berlekamp random coding bound \cite{berlekamp:bound}
\begin{align*}
{P^{(\mathsf B)}_{B}(l,k,\epsilon)}&{= \sum_{e=l-k+1}^l {l \choose e} \epsilon^e (1-\epsilon)^{l-e}} \\ &{+\sum_{e=1}^{l-k} {l \choose e} \epsilon^e (1-\epsilon)^{l-e}2^{-(l-k-e)}}
\end{align*}
which is an upper bound on the average block error probability of random codes, and the Singleton bound, which provides the block error probability of \ac{MDS} codes. The markers represent the results of Monte Carlo simulations. In order to obtain average results for the ensemble, the block error probability was averaged over $1000$ different {\ac{LRFC}} realizations. The bound in \eqref{eq:bound_Gavg} is very tight, as expected. Results for three different rates are shown in the figure. The highest rate corresponds to the use of the Hamming code alone, and the other two rates are $r=0.8$ and $r=0.5$. While for the Hamming code the performance lies in between the one of random codes and the one of \ac{MDS} codes, as the code rate decreases the performance of the scheme gets closer to the Berlekamp random coding bound, which means that for low rates our scheme performs almost as a random code.

\end{document}